\documentclass[a4paper,11pt]{article}
\usepackage{pos}
\usepackage[inline]{enumitem}
\usepackage{color}
\definecolor{cRed}{RGB}{196,0,100}
\newcommand{\modified}[1]{{{#1}}}

\title{Lifetime of the $B^+_c$ meson in relation to flavour anomalies}
\ShortTitle{Lifetime of the $B^+_c$ meson}

\author*[a]{Benjam\'\i{}n Grinstein}
\author[b]{Jason Aebischer}

\affiliation[a]{Department of Physics, UC San Diego\\
 9500 Gilman Dr, La Jolla, CA, USA}

\affiliation[b]{Physik-Institut, Universit\"at Z\"urich\\
  CH-8057 Z\"urich, Switzerland}

\emailAdd{bgrinstein@ucsd.edu}
\emailAdd{jason.aebischer@physik.uzh.ch}

\abstract{The Standard Model decay rate of the $B_c$ meson is
  discussed together with a novel approach that uses  experimental data in combination with an operator product expansion. In the new method differences of $B,\,D$ and $B_c$ meson decay rates are considered for which the free-quark contributions drop out, leading to a reduction of the theory prediction.}

\FullConference{%
  The Tenth Annual Conference on Large Hadron Collider Physics - LHCP2022\\
  16-20 May 2022\\
  online
}

\begin{document}
\maketitle

\section{Introduction}

The $B_c$ decay rate places stringent bounds on New Physics models in
the context of the $R(D)$ and $R(D^*)$ anomalies
\cite{Alonso:2016oyd,Blanke:2018yud,Blanke:2019qrx}, from
scalar Leptoquarks and Two-Higgs-Doublet models for example.

The $B_c=(\overline b c)$ meson is made up of two different heavy
quarks. It's decay is calculated using Non-Relativistic QCD (NRQCD),
expanding in  the small velocities of the two non-relativistic
quarks. After matching the relevant QCD operators onto
the NRQCD Lagrangian by integrating out the anti-particles of the
respective quarks, this approach allows to perform a systematic
expansion of the resulting NRQCD operators, which can be
carried out up to any given order and the truncation uncertainty
can be estimated. Furthermore, the symmetry properties of NRQCD allow
to relate the relevant matrix elements of the four-quark operators to
two parameters.

For the calculation we use the Optical Theorem to apply an Operator
Product expansion (OPE) performed on the forward scattering matrix
element of the $B_c$ meson. This OPE approach
\cite{Beneke:1996xe,Bigi:1995fs,Chang:2000ac} leads to similar results
as those obtained using QCD Sum Rules \cite{Kiselev:2000pp} or
Potential models \cite{Gershtein:1994jw}, which give lifetimes close
to the experimental value.  Experimentally, the $B_c$ decay rate has
been determined with rather small uncertainties by the LHCb
\cite{LHCb:2014ilr,LHCb:2014glo} and CMS \cite{CMS:2017ygm}
collaborations, yielding  an average of
\begin{equation}\label{eq:Gexp}
  \Gamma_{B_c}^\text{exp} = 1.961(35) \,\text{ps}^{-1}\,.
\end{equation}

In the following we will summarize the results from an updated OPE
computation of the $B_c$ decay rate
\cite{Aebischer:2021ilm,Aebischer:2022ptj}. We then outline a new
method to obtain $\Gamma_{B_c}$, using experimental results as well as a
non-perturbative expansions of the $B_c$, $B$ and $D$ mesons' lifetime.

\section{Results}

In Ref.~\cite{Aebischer:2021ilm}, the $B_c$ decay width was calculated
in the $\overline{\text{MS}}$, meson and Upsilon schemes. Neglecting
the strange-quark mass the result there is:
\begin{equation}\label{eq:ms0}
\begin{aligned}
  \Gamma^{\overline{\text{MS}}}_{B_c} &= (1.58\pm 0.40|^{\mu}\pm
  0.08|^{\text{n.p.}}\pm 0.02|^{\overline{m}}\pm 0.01|^{V_{cb}})\,\,\text{ps}^{-1}\,, \\
  \Gamma^{\text{meson}}_{B_c} &= ( 1.77\pm 0.25|^{\mu}\pm 0.20|^{\text{n.p.}} \pm 0.01|^{V_{cb}})\,\,\text{ps}^{-1} \,, \\
  \Gamma^{\text{Upsilon}}_{B_c} &= (2.51\pm 0.19|^{\mu}\pm 0.21|^{\text{n.p.}}\pm 0.01|^{V_{cb}})\,\,\text{ps}^{-1} \,.
\end{aligned}
\end{equation}

Here the main uncertainties result from the scale dependence,
indicated by $\mu$ in the above equation. It can be reduced by
including higher-order QCD corrections to the free-quark decay rates,
as well as to the Wilson coefficients involved. The second largest
uncertainty, indicated by $n.p.$ in eq.~\eqref{eq:ms0}, stems from
neglected higher-order corrections in the NRQCD expansion as well as
from uncertainties of the non-perturbative parameters. In the
$\overline{\text{MS}}$ scheme a non-negligible part of the uncertainty
results from the $\overline{\text{MS}}$ masses of the $\overline b$-
and $c$ quarks. Finally, there is an uncertainty due to the input
parameters, which is dominated by $V_{cb}$.

When including the strange-quark mass in the calculation for the free $c$-quark decays, the central values of the decay rate is reduced by about 7\%. We obtain in the three different schemes

\begin{equation}\label{eq:msnon0}
\begin{aligned}
  \Gamma^{\overline{\text{MS}}}_{B_c} &= (1.51\pm 0.38|^{\mu}\pm 0.08|^{\text{n.p.}}\pm 0.02|^{\overline{m}}  \pm0.01|^{m_s}\pm 0.01|^{V_{cb}})\,\,\text{ps}^{-1}\,, \\
  \Gamma^{\text{meson}}_{B_c} &= (1.70\pm 0.24|^{\mu}\pm 0.20|^{\text{n.p.}} \pm0.01|^{m_s}\pm 0.01|^{V_{cb}})\,\,\text{ps}^{-1} \,,  \\
  \Gamma^{\text{Upsilon}}_{B_c} &= (2.40\pm 0.19|^{\mu}\pm 0.21|^{\text{n.p.}} \pm0.01|^{m_s}\pm 0.01|^{V_{cb}})\,\,\text{ps}^{-1} \,.
\end{aligned}
\end{equation}
Besides the uncertainties mentioned above we have indicated the uncertainty due to $m_s$.

Within uncertainties the results in Eq.~\eqref{eq:msnon0} are
consistent with each other and with the experimental value in
Eq.~\eqref{eq:Gexp}. There is however a rather wide spread among the
three different mass schemes used. One strategy to improve on the
precision of the theory result is to reduce the uncertainty due to
scale-dependence. In the next section we discuss a novel approach
that reduces the dominant uncertainty, which as explained, is from the renormalization scale
dependence.

\section{Novel determination of $\Gamma_{B_c}$}\label{sec:newmethod}

The main theory uncertainty is mainly  due  to the free-quark decay rate, which is the leading term in the non-perturbative expansion of the decay rate of a meson $H_Q$ with heavy quark $Q$:
\begin{equation}\label{eq:GM}
  \Gamma(H_Q) = \Gamma_Q^{(0)}+\Gamma^{n.p.}(H_Q)+\Gamma^{\text{WA}+\text{PI}}(H_Q)+\mathcal{O}(\frac{1}{m_Q^4})\,,
\end{equation}
%
where the second term includes non-perturbative corrections and the
third term contains Weak Annihilation and Pauli Interference
contributions. The expansion in Eq.~\eqref{eq:GM} can be carried out
not only for the $B_c$ meson, but also for the $B$ and $D$
mesons. Taking now the difference of the three different decay rates
leads to:
\begin{align}\label{eq:diff}
  \Gamma(B)+\Gamma(D)-\Gamma(B_c) &= \Gamma^{n.p.}(B)+\Gamma^{n.p.}(D)-\Gamma^{n.p.}(B_c) \nonumber \\
  &+\,\Gamma^{\text{WA}+\text{PI}}(B)+\Gamma^{\text{WA}+\text{PI}}(D)-\Gamma^{\text{WA}+\text{PI}}(B_c)\,.
\end{align}
%
Since the free quark decay rate is independent of the meson state, it
drops out on the right-hand side of Eq.~\eqref{eq:diff}, thereby
reducing the uncertainty due to scale-dependence. For the computation
of $\Gamma(B_c)$, the decay rates of the $B$ and $D$ mesons can be
taken from experiment, whereas the right-hand side can be computed
using non-perturbative methods. The computation can be carried out for
charged or neutral $B$ and $D$ mesons, leading in principle to four
different ways to compute $\Gamma(B_c)$. In Tab.~\ref{tab:res} we show
the results for the $B_c$ decay rate in the meson \modified{and $\overline{\text{MS}}$} scheme, obtained
using the four different channels \cite{Aebischer:2021eio}.

\begin{table}[t]
\centering
 \begin{tabular}{|l |c |c |c |c|}
 \hline
 & $B^0,D^0$ & $B^+,D^0$ & $B^0,D^+$ & $B^+,D^+$ \\ [0.5ex]
 \hline \hline
   $\Gamma^{\text{meson}}_{B_c}$& 3.03 $\pm$ 0.54 & 3.04 $\pm$ 0.54 & 3.38 $\pm$ 0.98 & 3.39 $\pm$ 0.99 \\
  \hline
$\Gamma^{\overline{\text{MS}}}_{B_c}$ & 2.97 $\pm$ 0.42 & 2.98 $\pm$ 0.40 & 3.19 $\pm$ 0.80 & 3.19 $\pm$ 0.82
\\
 \hline
 \end{tabular}
 \caption{\small
Results obtained using the novel approach discussed in sec.~\ref{sec:newmethod} in the meson \modified{and $\overline{\text{MS}}$} scheme, using four different combinations of $B$ and $D$ mesons.
}
  \label{tab:res}
\end{table}

The results from this novel approach are in tension with the experimental result in eq.~\eqref{eq:Gexp}. Several reasons can be put forward to explain this disparity:
\begin{enumerate*}
  \item The uncertainties from NLO corrections to Wilson coefficients and free quark decay rates might be underestimated;
  \item Eye-graph contributions, neglected in lattice computations of matrix elements that we use~\cite{Becirevic:2001fy}, but estimated to be small using HQET sum rules \cite{King:2021jsq};
  \item Unexpectedly large contributions from higher dimension operators in the $1/m_Q$ expansion~\cite{King:2021xqp};
  \item Violation of quark-hadron duality.
\end{enumerate*}

A thorough analysis of the above mentioned points is in order to determine the reason for the discrepancy between the results and experiment.

\section{Summary}\label{sec:summary}

We have outlined the OPE approach to determine the $B_c$ decay rate in
the Standard Model. The results in the $\overline{\text{MS}}$, the
meson and the Upsilon scheme are compatible with each other and with
the experimental value. There is however a wide spread among the
central values in the three different mass schemes, where the main
uncertainties arise from neglected NLO QCD corrections.

We discussed a novel method to determine $\Gamma_{B_c}$ based on
differences of $B,\,D$ and $B_c$ decay rates that allows to reduce the
scale-dependence uncertainty. The results deviate significantly from
the experimental value, and we presented various possible reasons for
this discrepancy.

\acknowledgments
Work of BG  supported in part by the U.S. Department of Energy
Grant No.~DE-SC0009919.


\end{document}